\documentclass[aps,pre,twocolumn,groupedaddress,showpacs,superscriptaddress,amssymb,amsmath,longbibliography]{revtex4-2}
\usepackage{graphicx}
\usepackage{tabularx}
\usepackage{color}
\usepackage{amsmath}
\usepackage{comment}
\newcommand{\be}{\begin{equation}}
\newcommand{\ee}{\end{equation}}
\newcommand{\bea}{\begin{eqnarray}}
\newcommand{\eea}{\end{eqnarray}}
\usepackage{dcolumn}
\usepackage{hyperref}
\usepackage{bm}
\usepackage{epsf}
\usepackage{subfigure}
\usepackage{epstopdf}%
\usepackage{amsfonts}

\let\la\langle \let\ra\rangle
\let\bs\boldsymbol
\let\sf\mathsf

\newcommand{\Tr}{\textrm{Tr}}

\begin{document}

\title{Understanding multiple timescales in quantum dissipative dynamics: Insights from quantum trajectories}

\author{Matthew Gerry}
\affiliation{Department of Physics, University of Toronto, 60 Saint George St., Toronto, Ontario M5S 1A7, Canada}

\author{Michael J. Kewming}
\affiliation{School of Physics, Trinity College Dublin, College Green, Dublin 2, Ireland}

\author{Dvira Segal}
\affiliation{Department of Physics, University of Toronto, 60 Saint George St., Toronto, Ontario M5S 1A7, Canada}
\affiliation{Chemical Physics Theory Group, Department of Chemistry and Centre for Quantum Information and Quantum Control,
University of Toronto, 80 Saint George St., Toronto, Ontario M5S 3H6, Canada}
\email{dvira.segal@utoronto.ca}
\date{\today}

\begin{abstract}
Open quantum systems with nearly degenerate energy levels have been shown to exhibit long-lived metastable states in the approach to equilibrium, even when modelled with certain Lindblad-form quantum master equations. This is a result of dramatic separation of timescales due to differences between Liouvillian eigenvalues. These metastable states often have nonzero coherences which die off only in the long time limit once the system reaches thermal equilibrium.
We examine two distinct situations that give rise to this effect: one in which dissipative dynamics couple together states only within a nearly degenerate subspace, and one in which they give rise to jumps over finite energy splittings, between {\it separate} nearly degenerate subspaces.
We find, in each case, that a change of basis can often lead to a representation which more naturally captures the impact of the system-bath interaction than does the energy eigenbasis, revealing that separate timescales are associated with separate processes (e.g. decoherence into a non-energy eigenbasis, decay of population correlations to the initial state). 
This approach is paired with the inspection of quantum trajectories, which further provide intuition as to how open system evolution is characterized when coherent oscillations, thermal relaxation, and decoherence all occur simultaneously.
\end{abstract}

\maketitle
\section{Introduction}
Open quantum systems have been observed to exhibit, in some cases, a dramatic separation of timescales while relaxing towards equilibrium. Formally, this is attributed to significant differences in the magnitudes of different Liouvillian eigenvalues \cite{merkli2015,ivander_hyperacceleration_2023}. The phenomenon has received much attention in the context of anomalous relaxation, or the ``quantum Mpemba effect", which aims to speed up the approach to equilibrium by reducing or eliminating the slowest timescales from the evolution \cite{klich2019, carollo2021, manikandan2021, kochsiek2022}. 
In other studies, it has been considered whether intermediate metastable states that arise due to this timescale separation can offer advantages in applications such as quantum thermometry \cite{nick-thermometry}. Generically, the long-lived metastable state that is reached in the relaxation towards equilibrium exhibits coherences between energy eigenstates, which die off only over the longest timescale of the dynamics as the system reaches its thermal state \cite{tscherbul_long-lived_2014,tscherbul2015,dodin_secular_2018}.

The question of how coherence may be maintained over long time periods in open quantum systems has garnered much interest in recent years, particularly as many hope for advances in the field of quantum information processing, where it will be crucial to be able to do so \cite{mikeandike}. The significance of coherences has also been investigated in the context of quantum transport and quantum thermodynamics, where it has been questioned whether they can lead to improved performance of thermal machines \cite{scully2011, michaelQAR, hammam2021, um2022}, or play a role in the violation of cost-precision tradeoff relations \cite{bijayTUR, junjieTUR, turzero}. Other studies have considered the role of coherence in biological processes, including photosynthesis \cite{engel2007, kassal2013} and vision \cite{prokhorenko2006, tscherbul2015_retinal}.

One set of research directions towards this goal comprises describing the time evolution of open quantum systems in a manner that accurately captures the effects of coherence. For open systems weakly coupled to thermal baths, the time evolution is often described using quantum master equations (QMEs), obtained by starting from unitary evolution of the full system plus baths, making the Born-Markov approximation, and tracing over environmental degrees of freedom to obtain an expression for the time derivative of the reduced density operator of the system \cite{breuer-book}. The resulting Redfield equation captures the effects of coherence in the sense of interest to us here, however it is known not to be of Gorini-Kassakowski-Lindblad-Sudarshan (GKLS) form \cite{gorini1976,lindblad1976}. It therefore fails, in general, to preserve the positivity of the system reduced density operator \cite{kohen1997,hartmann2020}. In addition, the Redfield equation does not guarantee the satisfaction of fluctuation symmetries, leading in some cases to violations of the second law of thermodynamics \cite{argentieri2014, vmodel_fluc_sym}. These pitfalls are often addressed by approximating further, and using the fully Secular, or Davies, Lindblad-form QME which dispenses with oscillating terms in the Redfield equation \cite{breuer-book,davies1974}. For systems lacking strict degeneracies, however, this equation neglects any impact that coherences between energy eigenstates can have on population evolution. It is thus inadvisable to make the full secular approximation when describing systems with eigenstates close in energy, for which some terms in the Redfield equation oscillate on timescales slow enough to rival those of the dynamics of interest \cite{trushechkin_unified_2021}.

GKLS-form master equations that are obtained without making the full secular approximation have been derived to describe dynamics more accurately in the presence of nearly degenerate levels \cite{trushechkin_unified_2021,mccauley_accurate_2020,potts_thermodynamically_2021,farina2019}. The so-called ``unified" quantum master equation (UQME), which shall be our focus, is obtained from the Redfield equation by removing rapidly oscillating terms but keeping slowly oscillating terms--that is, those terms oscillating at frequencies set by the energy differences between pairs of nearly degenerate states. GKLS form is achieved by neglecting these small energy splittings in the calculation of bath-induced transition rates \cite{trushechkin_unified_2021}. This master equation has been shown not only to preserve the positivity of the reduced density operator, but also to satisfy fluctuation symmetry with respect to heat transport away from equilibrium, thus giving dynamics that satisfy the second law of thermodynamics \cite{vmodel_fluc_sym,soret2022}. When pairs of nearly degenerate states are present in the system, this equation gives results for the elements of the reduced density operator much closer in value to the predictions of the Redfield equation than does the fully secular QME.

For open quantum systems with sets of nearly degenerate levels--those for which long-lived coherences tend to arise during the relaxation to equilibrium--there are, broadly, two situations of interest. Each can arise in isolation or both simultaneously, as determined by the form of the system-bath interaction. In the first, the dissipative dynamics couple together states within a nearly degenerate subspace. Such dynamics cause no change to the expectation value of the system Hamiltonian larger than the order of the small energy splitting. A simple example of such a situation would be a nearly degenerate two-level system coupled to a thermal bath in a manner that leads to transitions that build coherence between the levels.

In the second situation, dissipative dynamics lead the system to jump \textit{between} distinct nearly degenerate subspaces, such that the jumps themselves are associated with a finite energy splitting and the timescales for relaxation generally have explicit temperature dependence. This latter case is exemplified by the ``V" model, which has a pair of nearly degenerate excited states separated by a finite energy gap to a ground state. This model has been used as a paradigm for certain atomic systems in quantum optical studies \cite{hegerfeldt1992,hegerfeldt1993}, as well as nitrogen vacancy centres \cite{hernandez-gomez2022}. When studied using the Redfield equation, it has been found to exhibit dramatically distinct timescales during relaxation towards equilibrium, including long-lived coherences in the metastable state \cite{tscherbul_long-lived_2014,dodin_quantum_2016, dodin_secular_2018}. It has also shown coherences at steady state in nonequilibrium situations \cite{li2015,ivander_quantum_2022}. These effects are captured adequately by the UQME, both in numerical simulations and approximate methods for obtaining analytic solutions to the master equation \cite{vmodel_fluc_sym, ivander_quantum_2022, ivander_hyperacceleration_2023}. The separation of timescales in the dynamics is attributed to the fact that the different eigenvalues of the Liouvillian describing the system's time evolution span multiple orders of magnitude \cite{ivander_hyperacceleration_2023}, however, intuition as to how the interaction with a thermal bath can facilitate the existence of these long-lived coherence is lacking.


Gaining intuition in this context involves studying systems with nearly degenerate states through quantum trajectories.
Quantum trajectories, a concept with a longstanding history in quantum optics, were formally developed to depict quantum systems undergoing continuous measurements \cite{gardiner1992, wiseman-book, Jacobs2014, landi_current_2023}. In this scenario, the measurement process results in the stochastic evolution of a given trajectory. Averaging over many such trajectories recovers the ensemble average, described by a GKLS-form master equation.

To be more precise, a quantum trajectory is obtained by ``unraveling" the master equation: examining a specific exemplary sequence of ``jumps" and describing the evolution of the reduced density operator conditioned on the sequence $\rho_c$. This evolution occurs discontinuously at random instants when jumps happen. If the interaction with the environment involves continuous measurement (e.g., a detector in the environment clicks each time a jump occurs), $\rho_c$ can be understood as the system's state conditioned upon a particular sequence of clicks. Taking the ensemble average of conditional density operators over the possible sequences then recovers the system's reduced density operator, $\rho$.




In this study, we consider minimal models that exhibit dramatically different timescales and long-lived coherences during their relaxation towards equilibrium. Our objective is to root these phenomena in underlying physical processes. Using the UQME--a natural choice of GKLS master equation which captures the timescale effects--we investigate the dynamics of these systems as they thermalize. We build insight by carrying out the analysis not only in the energy eigenbases of the systems considered, but also in alternate bases that more naturally reflect the way the system-bath interaction manifests in the evolution of the state. 
This approach also lends itself elegantly to the unravelling of the master equation to observe quantum trajectories, which, in turn, help to further paint the picture of why separate timescales emerge in the overall dynamics. We do note, however, that the measurement operators do not constitute a clearly defined measurement process, but rather that the unravelled dynamics can shed light onto the separation of time scales.


In Sec. \ref{sec:2ls}, we consider 
the case where dissipative dynamics act within a nearly degenerate subspace and work to build intuition for the physical processes underlying dramatic timescale separation during the relaxation to equilibrium. In Sec. \ref{sec:V} we turn our attention to the case in which dissipative dynamics act between distinct nearly degenerate subspaces, and identify and explain analogous timescale separation. In Sec. \ref{sec:summary} we summarize our findings and discuss how they are situated amongst a broader set of questions.

\section{Dissipative dynamics within a nearly degenerate subspace}\label{sec:2ls}

\begin{figure}
    \centering
    \includegraphics[width=0.65\columnwidth, trim = 10 10 10 10]{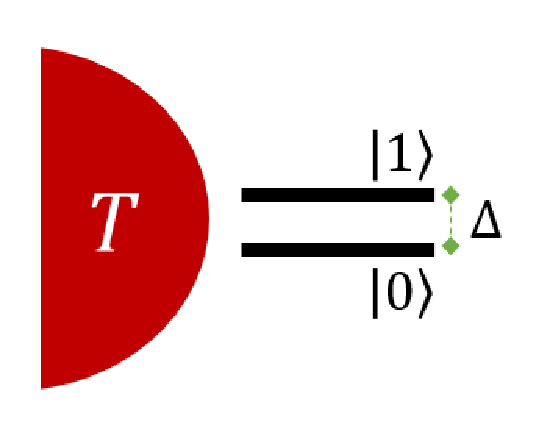}
    \caption{Level diagram for the two-level system with splitting $\Delta$, coupled to a bosonic bath at temperature $T$ via the system-bath part, $H_{SB}$, of the Hamiltonian.}
    \label{fig:2ls_diagram}
\end{figure}

\subsection{Setup}\label{sec:2ls_setup}
In order to probe 
the case in which dramatically different relaxation timescales occur as a result of dissipative dynamics acting within a nearly degenerate subspace, we consider a basic model consisting of a spin (2-level system) coupled weakly to a reservoir. The full system plus environment is described by the Hamiltonian,
\be 
    H = H_S + H_B + H_{SB}.
\ee
The system Hamiltonian is 
\be 
    H_S = \frac{\Delta}{2}\left(|1\ra\la1|-|0\ra\la0|\right).
\ee
The level diagram for this model, along with the bath, is depicted in Fig. \ref{fig:2ls_diagram}.

$H_{SB}$ describes the coupling between the system and bath, and is of the simple form
\be\label{eq:V_2ls}
    H_{SB} = S\otimes B
\ee
where $S$ is an operator acting solely on the system Hilbert space, and $B$, on the bath.

Crucially, the `system part' of the interaction Hamiltonian is given by
\be 
    S = \frac{1}{\sqrt{2}}\left(|0\ra\la0| - |1\ra\la1| + |0\ra\la1| + |1\ra\la0|\right).
\ee
In the energy eigenbasis, this is represented by a matrix $\bs{\sf{S}}=\left(\sigma_x + \sigma_z\right)/\sqrt{2}$.

We wish to derive the UQME for this system, which consists of transition rates whose specific values depend on the properties of the bath. These are obtained by evaluating the rates of the Redfield equation, which are, in general, functions of two distinct frequencies, at only the average of the two arguments. This captures the assumption of near degeneracy, effectively neglecting the difference between two energy splittings and reduces the rate to a function of a single frequency \cite{trushechkin_unified_2021}, i.e.,
\be\label{eq:red_to_uni}
    \gamma(\omega_1,\omega_2)\rightarrow \gamma(\omega)\equiv\gamma(\omega,\omega),
\ee
where $\omega\equiv(\omega_1+\omega_2)/2$. Such rates can be determined for bosonic or fermionic baths, and for any form of the spectral density. To probe the phenomenon of dramatic timescale separation for this two-level system, we require only one additional assumption: that we work in the limit of $\Delta\ll T$. That is, the two levels are nearly degenerate relative to the energy scale associated with the temperature, justifying the approximation described by Eq~(\ref{eq:red_to_uni}).

In this case, the only transition rate describing the dissipative dynamics under the UQME is the one where the argument goes to zero:
\be\label{eq:rates} 
    \gamma\equiv\lim_{\omega\rightarrow0}\gamma(\omega)
\ee
The lack of any large energy splittings in the system means that no terms in the Redfield equation are discarded to obtain the unified QME. The UQME is obtained from Redfield simply by taking all transition rates to the approximate value $\gamma$. In the Schrodinger picture, the UQME is given by
\be\label{eq:uqme_2ls}
    \frac{d\rho}{dt} = \mathcal{L}\rho = -i[H_S,\rho] + \mathcal{D}[L]\rho,
\ee
where $\rho\equiv\Tr_B[\rho_{tot}]$ is the system reduced density operator, and,
\be\label{eq:dissipator}
    \mathcal{D}[L]\rho = L\rho L^\dag - \frac{1}{2}\{L^\dag L,\rho\}.
\ee
There is only one dissipator corresponding to a single collapse, or ``Lindblad", operator, which is directly proportional to the system part of the interaction Hamiltonian:
\be 
    L = \sqrt{\gamma}S.
\ee

While it is clear that Eq.~(\ref{eq:uqme_2ls}) is of Lindblad form, since the single positive value $\gamma$ clearly amounts to a positive semidefinite ``matrix", this is in contrast to typical uses of fully secular quantum master equations, where there are often many collapse operators, and they tend of be of the form of projectors onto, or jumps between, energy eigenstates. 
Nonetheless, Eq.~(\ref{eq:uqme_2ls}) describes evolution of the density operator for an ensemble.

The evolution at the trajectory level, however, is described by a stochastic jump equation \cite{wiseman-book},
\begin{equation}
\label{eq:trajectory}
    d \rho_{c} = -\mathcal{H}
    \left[i H_{S} + \frac{1}{2}L^{\dagger}L\right]\rho_{c}dt +  dN(t)\mathcal{G}\left[L\right]\rho_{c}\,,
\end{equation}
where $dN(t)$ is a stochastic Poisson increment satisfying $dN(t)^{2} = dN(t)$ and $\rho_{c}$ is the conditioned density matrix.
The two superoperators are defined by
\begin{equation}
    \mathcal{G}[L]\rho = \frac{L\rho L^{\dagger}}{{\rm Tr}[ L\rho_{c}L^{\dagger} ]} - \rho\,, \quad 
    \mathcal{H}[L]\rho = L\rho + \rho L^{\dagger} - {\rm Tr}[ \hat{L}\hat{\rho} + \hat{\rho}\hat{L}^{\dagger} ]\,.
\end{equation}
Each detection event corresponds to a quantum jump given by the superoperator $\mathcal{G}[L]$. 
The average jump rate is given by ${\rm E}[dN(t)]={\rm Tr}[L \rho L^\dagger]dt$, where ${\rm E}[X]$ denotes a classical average over stochastic trajectories.
Thus, when $dN(t) =0$, the evolution of the system will be governed by the ``null-meassurement'' term---corresponding to the first term in Eq.~(\ref{eq:trajectory}). However, when a jump occurs the state is instantly updated according to the projection onto $L$ as determined by the second term. 

While this assumption is by no means necessary, we consider, for the purpose of calculations, a bosonic bath, with
\be\label{eq:HB_2ls}
    H_B = \sum_k\omega_kb_{k}^\dag b_k,
\ee
where $b_k$ is the annihilation operator for a mode $k$ with energy $\omega_k$. The bath is taken to be at thermal equilibrium with temperature $T$, so the number operators satisfy $\langle b_k^\dag b_k\rangle=n(\omega_k)$, with $n(\omega)=(e^{\beta\omega}-1)^{-1}$, the Bose-Einstein distribution at inverse temperature $\beta\equiv1/T$ (we take the Boltzmann constant, $k_B=1$). As such, the transition rate appearing in the UQME is given by
\be\label{eq:rates}
    \gamma = \lim_{\omega\rightarrow0}n(\omega)\mathcal{J}(\omega)
\ee
where $\mathcal{J}(\omega)$ is the spectral density of the bath. Given an Ohmic bath, $\mathcal{J}(\omega)=a\omega$, we have $\gamma=aT$.

\subsection{Energy Eigenbasis}\label{sec:2ls_eigenbasis}
We can write the equations of motion for the distinct elements of $\rho$ in the energy eigenbasis as given by the unified quantum master equation. It is convenient to define the ``polarization", $P=(\rho_{11}-\rho_{00})/2$. We use the superscripts $R$ and $I$ to denote the real and imaginary parts of complex-valued matrix elements. Utilizing the condition of normalization to reduce the system down to three equations, we have
\begin{align}\label{eq:uqme_elements}
    \dot{P} &= -\gamma(P + \rho_{10}^R)
    \nonumber\\
    \dot{\rho}_{10}^R &= -\gamma(P + \rho_{10}^R) + \Delta \rho_{10}^I
    \nonumber\\
    \dot{\rho}_{10}^I &= -\Delta \rho_{10}^R - 2\gamma \rho_{10}^I.
\end{align}

\begin{figure}
    \centering
    \includegraphics[width=0.75\columnwidth, trim = 20 10 20 10]{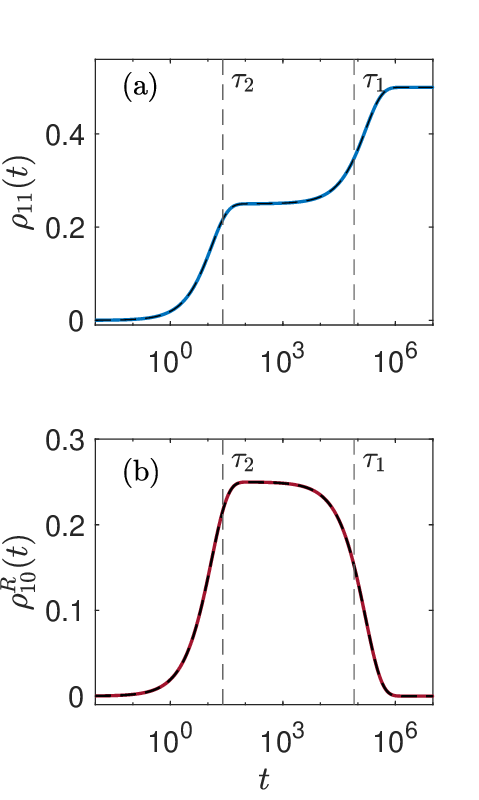}
    \caption{(a) The excited state population and (b) the real part of the eigenstate coherence as a function of time, for a two-level system initially in the ground state, Eqs. (\ref{eq:uqme_elements})-(\ref{eq:2LS_rho_eigen}). Analytic results from the perturbative method (dashed) are plotted along with numerical values obtained using the full Unified QME (solid), showing good agreement. The two characteristic timescales, $\tau_1$ and $\tau_2$, describing the stages of relaxation towards equilibrium are shown as vertical dashed lines. Calculations are carried out with $\Delta=0.001$, $T=1$, and $a=0.02$, with $a$ the dimensionless coupling coefficient as defined below Eq.~(\ref{eq:rates}).}
    \label{fig:rho_2ls}
\end{figure}

These are the equations of motion consistent with using the UQME, and give results which show good agreement with full Redfield in the case where nearly degenerate levels are present \cite{vmodel_fluc_sym}. However, other choices of how to truncate the Redfield equation have been employed in studies of similar models that emphasize other phenomena, such as coherences that may arise at steady state \cite{guarnieri2018,purkayastha_tunable_2020}. Critically, any choice that eliminates the coupling of populations to coherences in the energy eigenbasis, as is seen here, will not give rise to sets of dramatically different timescales in the approach to equilibrium.

Given a choice of initial state, we can obtain analytic expressions for elements of $\rho$ as a function of time by solving for the eigevalues of the Liouvillian up to leading order in $\Delta$. Following the approach of Ref. \cite{ivander_hyperacceleration_2023}, we treat the $\Delta$-dependent part of $\mathcal{L}$ as a perturbation, first solving for the eigenvalues of the ``unperturbed" Liouvillian and then and obtaining the order-$\Delta$ correction only for the eigenvalue that vanishes in the $\Delta\rightarrow0$ limit. Supposing the system is initially in its ground state, $\rho(0)=|0\ra\la0|$, we obtain the following expressions for the population of the excited state, $\rho_{11} = P+1/2$, and the real part of the coherence, $\rho_{10}^R$, as functions of time:
\begin{align}\label{eq:2LS_rho_eigen}
    \rho_{11}(t) &= \frac{1}{2}-\frac{1}{4}\left( e^{-\frac{\Delta^2}{4\gamma}t} +e^{-2\gamma t} \right)
    \nonumber\\
    \rho_{10}^R(t) &= \frac{1}{4}\left( e^{-\frac{\Delta^2}{4\gamma}t} -e^{-2\gamma t}\right).
\end{align}
Note that our analytic approach is not sufficient to solve for $\rho_{10}^I(t)$ with this particular choice of initial state, since it is initially zero and thus, by Eq.~(\ref{eq:uqme_elements}), the leading order contribution to its evolution is of order $\Delta$. It is not valid to treat contributions of this order as perturbations.

Present in both expressions of Eq.~(\ref{eq:2LS_rho_eigen}) are two distinct timescales, $\tau_1=4\gamma/\Delta^2$ and $\tau_2=1/(2\gamma)$, corresponding to the reciprocals of the nonzero eigenvalues of the Liouvillian for this system, up to leading order in $\Delta$. This describes evolution in the relatively short term to a metastable state which exhibits nonzero coherence, before evolving to the true steady state of the system, which is the maximally mixed state $\rho_\infty=I_2/2$, where $I_2$ is the identity operator on a 2-dimensional space. This is also the expected thermal state for this system, once again neglecting order-$\Delta$ contributions. This behaviour is demonstrated in Fig. \ref{fig:rho_2ls}. Furthermore, while different choices of the initial state lead to different functional forms for the elements of $\rho$, the same timescales $\tau_1$ and $\tau_2$ arise independently of this choice, with the exception of trivial cases such as evolution beginning in the steady state. 



\subsection{Alternate basis}\label{sec:2ls_alternate}
We take the fact that both timescales, $\tau_1$ and $\tau_2$, show up in the expressions for both the populations and coherence in the energy eigenbasis as motivation to transform to a new basis in hopes of gaining some insight into the physical origin of separate timescales. Namely, we identify the unitary matrix $\bs{\sf{V}}$ which diagonalizes the Lindblad operator, i.e.,
\be
    \bs{\sf{L}} = \bs{\sf{V}}\Tilde{\bs{\sf{L}}}\bs{\sf{V}}^{-1},
    \label{eq:Ldiag}
\ee
where $\bs{\sf{L}}$ and $\Tilde{\bs{\sf{L}}}$ are the matrices representing $L$ in the energy eigenbasis and new basis, respectively, with the latter diagonal. We find $\bs{\sf{V}}$ to be given by
\be
    \bs{\sf{V}} = \frac{i}{\sqrt{2}}\left({\begin{array}{cc}
                                -(1-1/\sqrt{2})^{1/2} & (1+1/\sqrt{2})^{1/2} \\
                                \left[2(1-1/\sqrt{2})\right]^{-1/2} & \left[2(1+1/\sqrt{2})\right]^{-1/2}
                                \end{array}}\right),
\ee
and the collapse operator is simply represented as $\Tilde{\bs{\sf{L}}}=-\sqrt{\gamma}\sigma_z$.
We denote the new basis states $\{|\psi_+\ra, |\psi_-\ra\}$; they are related to the energy eigenstates via
\be
    |\psi_\pm\ra = \frac{i}{\sqrt{2}}\left(\pm\sqrt{1\pm\frac{1}{\sqrt{2}}}|0\ra + \frac{1}{\sqrt{2}}\frac{1}{\sqrt{1\pm\frac{1}{\sqrt{2}}}}|1\ra\right).
\ee
The system Hamiltonian is clearly not diagonal in this basis. It is expressed in terms of the new basis states as
\begin{align}
    H_S = \frac{1}{\sqrt{2}}\frac{\Delta}{2}\big(&|\psi_-\ra\la\psi_-| - |\psi_+\ra\la\psi_+|\nonumber\\
    & + |\psi_+\ra\la\psi_-| + |\psi_-\ra\la\psi_+|\big).
\end{align}

\begin{figure}
    \centering
    \includegraphics[width=0.75\columnwidth, trim = 20 10 20 10]{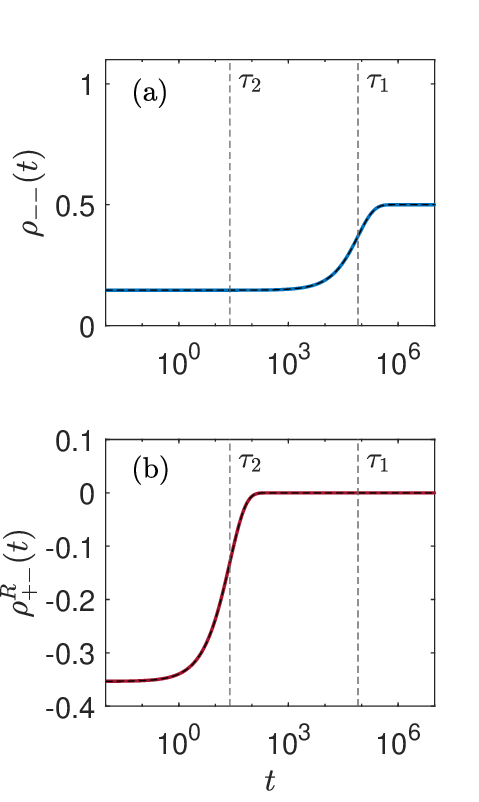}
    \caption{(a) The level population and (b) real part of the coherence in the alternate basis [Eqs. \ref{eq:uqme_2ls_alt}-\ref{eq:2LS_rho_alt}] as a function of time, starting from the ground state of the system Hamiltonian. The solid curves represent the results of solving the master equation numerically, while the dashed black curves are the approximate analytic expressions. The two timescales are shown as vertical dashed lines. Parameter values are the same as in Fig. \ref{fig:rho_2ls}.}
    \label{fig:rho_2ls_alternate}
\end{figure}


The benefit of working in this new basis is that it offers a straightforward physical interpretation of the effect of the dissipative part of the dynamics at the trajectory level. At random instances in time, characterized by the rate $\gamma$, a ``jump" occurs, taking the system to a new state. 
This transition amounts to flipping the signs of the off-diagonal elements of $\rho_c$ in the new basis, with no change to the diagonal elements. That is, the system-bath interaction can be understood as giving rise to pure decoherence. However, this decoherence is into a basis other than the energy eigenbasis, and is therefore paired with Rabi oscillations between basis states.

The fact that the system-bath Hamiltonian couples together states only within a nearly degenerate subspace is a requirement for the interaction to be interpreted in this way. A finite energy splitting would require distinct upward and downward transition rates, leading to different Lindblad operators which would not be simultaneously diagonalizable. The presence of just a single Lindblad operator proportional to the interaction Hamiltonian guarantees its diagonalizability, i.e., the existence of a basis in which it can be interpreted as describing pure decoherence \cite{breuer-book}. This implies that, while we have focused for our examples on a specific form for the system part, $S$, of $H_{SB}$, the relevant basis could be identified given any choice of Hermitian operator $S$, provided all else is held equal.

To determine how the state of our exemplary system evolves as a function of time under the same conditions as in Sec. \ref{sec:2ls_eigenbasis}, we obtain the initial state (ground state of $H_S$) in this new basis. As a matrix,
\be 
    \Tilde{\bs{\rho}}(0) = \bs{\sf{V}}^{-1}\bs{\rho}(0)\bs{\sf{V}} = \frac{1}{2}\left(I - \frac{1}{\sqrt{2}}\left(\sigma_x+\sigma_z\right)\right).
\ee
We obtain an expression for the top-left matrix element of $\Tilde{\bs{\rho}}$, which, due to the change of basis, represents the population not of the {\it ground} state, but rather, of the state $|\psi_-\ra$. We denote this $\rho_{--}$ and the population of state $|\psi_+\ra$, $\rho_{++}$. We also solve for the real and imaginary parts of the coherence between the new basis states, $\rho_{+-}\equiv\la\psi_+|\rho|\psi_-\ra$. This amounts to solving the UQME, which takes the form of the following set of equations, where we define the variable $\Tilde{P}\equiv(\rho_{--}-\rho_{++})/2$:
\begin{align}\label{eq:uqme_2ls_alt}
    \dot{\Tilde{P}} &= \frac{\Delta}{\sqrt{2}}\rho_{+-}^I\nonumber\\
    \dot{\rho}_{+-}^R &= -2\gamma\rho_{+-}^R -\frac{\Delta}{\sqrt{2}}\rho_{+-}^I\nonumber\\
    \dot{\rho}_{+-}^I &= -\frac{\Delta}{\sqrt{2}}\left(\Tilde{P}-\rho_{+-}^R\right) - 2\gamma\rho_{+-}^I.
\end{align}

Using the same perturbative method with which we solved for the matrix elements in the eigenbasis, we obtain expressions for $\rho_{--}$ and $\rho_{+-}^R$ as functions of time, transforming back to the density matrix elements from the solution for $\Tilde{P}$.
\begin{align}\label{eq:2LS_rho_alt}
    \rho_{--}(t) &= \frac{1}{2}-\frac{1}{2\sqrt{2}}e^{-\frac{\Delta^2}{4\gamma}t}
    \nonumber\\
    \rho_{+-}^R(t) &= -\frac{1}{2\sqrt{2}}e^{-2\gamma t}.
\end{align}
Once again, variations in $\rho_{+-}^I$ are of order $\Delta$ and smaller and thus are not captured by the analytic methods, so we focus on $\rho_{--}$ and $\rho_{+-}^R$ for the timescale analysis. In addition, as before, the maximally mixed state is approached in the long time limit, as expected due to its basis independence.

\begin{figure*}%
    \centering
    \includegraphics[width=0.43\textwidth, trim=20 10 10 0]{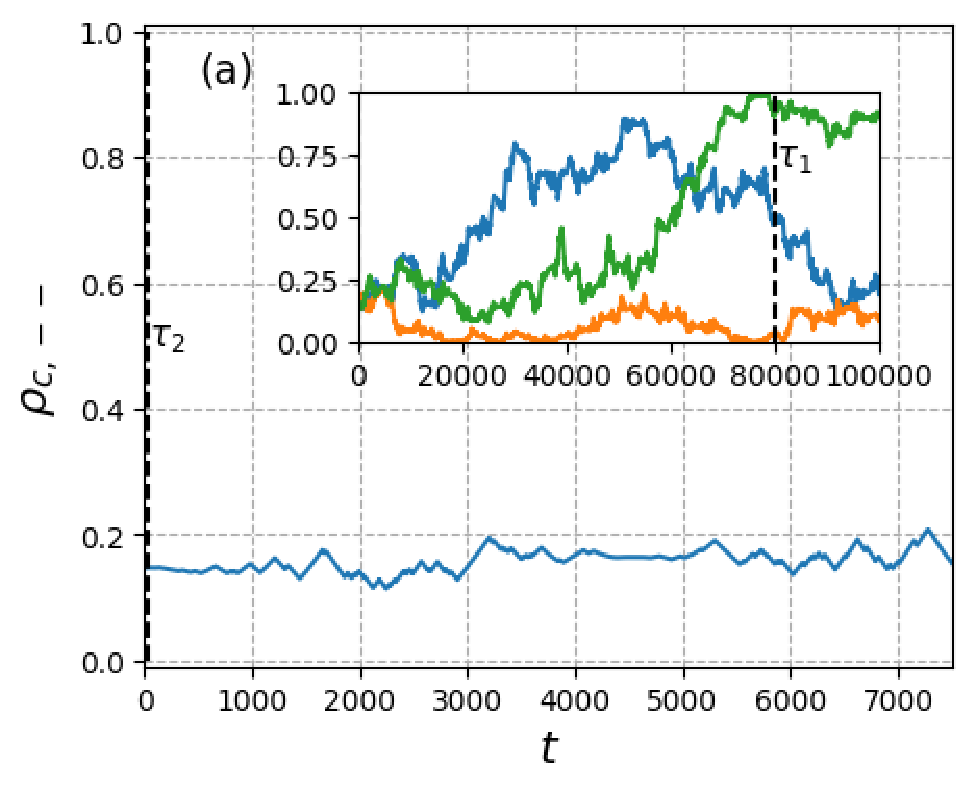}%
    \qquad
    \includegraphics[width=0.43\textwidth, trim=20 10 25 40]{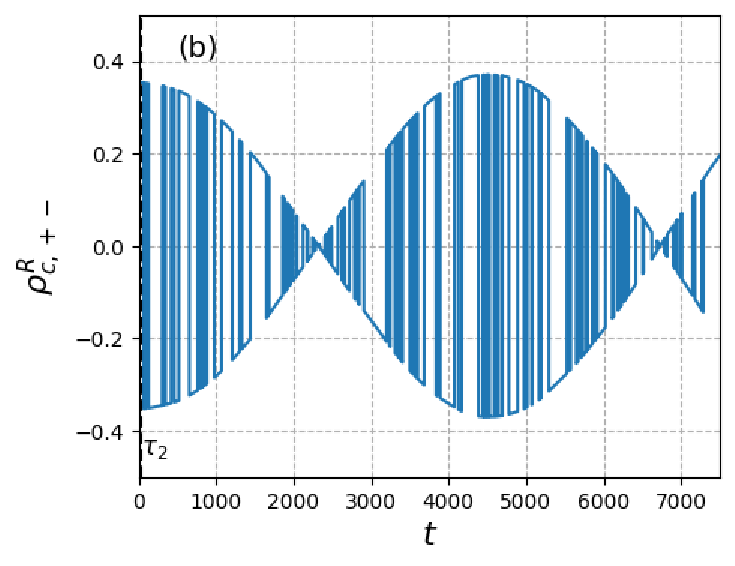}%
    \caption{A sample individual trajectory for (a) the population of $|-\ra$ and (b) the real part of the coherence under the evolution described in Section \ref{sec:2ls} in the alternate basis
    [Eqs. (\ref{eq:Ldiag})-(\ref{eq:2LS_rho_alt})].
    $\tau_2$ is displayed in each panel as a vertical dashed line, very close to the origin. Initial state and parameter values are the same as in Fig. \ref{fig:rho_2ls}. Inset of (a): the population evolution is shown over a longer time interval for three sample trajectories, demonstrating that all maintain similar values at short times. $\tau_1$ is shown as a vertical dashed line in the inset, by which time, correlations between the current and initial values are lost.}%
    \label{fig:2ls_traj}%
\end{figure*}

As in Eq.~(\ref{eq:2LS_rho_eigen}), we note the two distinct timescales, $\tau_1$ and $\tau_2$. However, rather than both appearing in both expressions, $\tau_1$ can be interpreted strictly as a timescale for the decay of populations in this basis towards their steady state values, while $\tau_2$ is strictly a timescale for the decay of coherences to zero. This is demonstrated in Fig. \ref{fig:rho_2ls_alternate}, wherein the evolution of each element of $\rho$ appears as simple exponential decay. It is still the case that the system first evolves to a quasi-steady state, where it remains while $\tau_2<t<\tau_1$. In this basis, this state is characterized simply by the coherence having decayed but the populations remaining close to their initial values. The values plotted in Fig. \ref{fig:rho_2ls_alternate} may be transformed back to the energy eigenbasis via the matrix $\bs{\sf{V}}$, to give results very similar to those obtained when solving the UQME directly in the energy eigenbasis.

\subsubsection{Trajectory Analysis}
To gain some insight as to why the unified QME in this alternate basis gives rise to the behaviour expressed by Eq.~(\ref{eq:2LS_rho_alt}) and shown in Fig. \ref{fig:rho_2ls_alternate}, we can look at an example trajectory in this basis for the system under the evolution described. 
This is generated using the QuTip python library \cite{johansson_qutip_2012,johansson_qutip_2013}, and shown in Fig. \ref{fig:2ls_traj}. Averaging over such trajectories recovers the time evolution of the density operator plotted in Fig. \ref{fig:rho_2ls_alternate} in the limit of an infinite sample size.

Since the system Hamiltonian is not diagonal in this basis, the unitary contribution to evolution under the UQME is characterized by Rabi oscillations with angular frequency $\Delta$. In the absence of any interaction with the bath, this leads to oscillations of the off-diagonal elements along the {\it envelope} in Figure \ref{fig:2ls_traj}(b), exactly out of phase with corresponding population oscillations. 
The interaction with the bath, however, causes the sign of the off-diagonal elements to flip at randomly timed jumps with rate $\gamma$. At the level of an ensemble, these sign-flipping events lead coherences to die off on the timescale $\tau_2$, as while the initial state dictates that all trajectories begin with $\rho_{c,+-}^R(0)=-1/(2\sqrt{2})$, positive and negative values that arise after varying sequences of jumps lead to a result of zero when averaging over many trajectories. Conversely, these jumps help the populations remain closer to their initial values for longer than they would under isolated Rabi oscillations.

To understand this, we consider the evolution of $\rho_c$ on the Bloch sphere, where the state is represented by a vector $\vec{s}$ with components $s_{x}=2\rho_{c,+-}^{R}$, $s_{y}=2\rho_{c,+-}^{I}$, and $s_z=\rho_{c,--}-\rho_{c,++}$. The coherent part of its time evolution, amounting to Rabi oscillations, is given by
\be\label{eq:rabi}
    \dot{\vec{s}} = \vec{\Omega}\times\vec{s},
\ee
where $\times$ represents the cross product in three dimensions and $\vec{\Omega}=(\Delta/\sqrt{2},0,\Delta/\sqrt{2})^T$ is the so-called ``drive vector". This is equivalent to the equations of motion, Eq.~(\ref{eq:uqme_2ls_alt}), but without the effects of dissipation. At the trajectory level, these effects are instead reflected in the random jumps that occur, over and above the evolution described by Eq.~(\ref{eq:rabi}).

\begin{figure*}
    \centering
    \includegraphics[width=0.9\textwidth, trim=50 10 50 10]{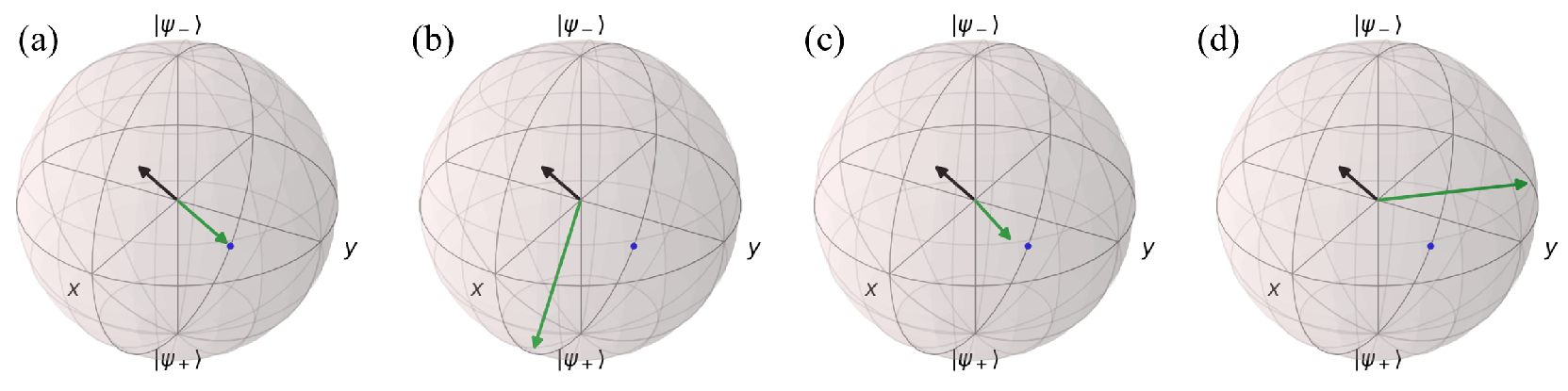}
    \caption{Various snapshots of the evolution of one sample trajectory, depicted as a vector on the Bloch sphere (green), where the x, y, and z axes represent the real and imaginary parts of the coherence in the alternate basis, and the population polarization between states $|-\ra$ and $|+\ra$, respectively. The initial state is shown throughout as a blue dot. The ``drive'' vector, $\vec{\Omega}$, about which the Bloch vector precesses in Rabi oscillations, is also shown (black). Note that this quantity has units of frequency, thus only its direction is meaningful as shown in these diagrams. (a) The initial state is the ground state, and is therefore stationary. (b) An instantaneous jump occurs after a time period of order $\tau_2$, represented by a reflection of the Bloch vector across the $z$-axis. There is now a finite angle between the Bloch vector and the drive vector, so Rabi oscillations proceed. (c) After another jump, the Bloch vector is slightly offset from its initial position, so the precession continues, albeit at a much slower rate. Jumps continue to occur at a rate set by the timescale $\tau_2$. (d) After a time period on the order of $\tau_1$, the Bloch vector is in the vicinity of the equator, indicating that, at the ensemble level, the system has reached the maximally mixed state, which corresponds to the true steady state for this system. Parameter values are the same as in Fig. \ref{fig:rho_2ls}.}
    \label{fig:bloch}
\end{figure*}

Our initial state corresponds to a vector in the lower-right quadrant of the $xz$-plane, strictly antiparallel to $\vec{\Omega}$, as shown in Fig. \ref{fig:bloch}(a). The state is stationary; it is, after all, the ground state of the system Hamiltonian. However, before long, a jump occurs, reflecting $\vec{s}$ across the $z$-axis (Figure \ref{fig:bloch}(b)). $\vec{s}$ is suddenly perpendicular to $\vec{\Omega}$, around which it begins to precess with angular frequency $\Delta$, bringing it out of the $xz$-plane into the $-y$ region. $\dot{s}_z\propto s_y$, so $s_z$ begins to decrease, but only until the next jump takes the state into the $+y$ region. After this point, $s_z$ begins to increase again, reversing the previous decrease (Figure \ref{fig:bloch}(c)). Since $\gamma\gg\Delta$, changes to $s_z$ (i.e., changes to the state {\it populations}) rarely have the opportunity to accumulate much before a jump occurs and they are undone. This is why, at short times, $\rho_{c,--}$ undergoes the jagged evolution around its initial value demonstrated in Fig. \ref{fig:2ls_traj}. Only on much longer timescales, on the order of $\tau_1$, does the Bloch vector reach the vicinity of the equator, as in Fig. \ref{fig:bloch}(d). This signifies that the correlations to the initial value of the populations have died off and the system is truly in a thermal state.

One may expect the coherences to exhibit similar behaviour, since it is also the case that $\dot{s}_x\propto s_y$. The key difference is that while a jump does, therefore, lead $\dot{s}_x$ to change sign, so simultaneously does $s_x$ {\it itself}. This allows changes to $|s_x|$ to accumulate seamlessly rather than being constantly undone, in a manner that effectively characterizes uninterrupted Rabi oscillations up to a frequently flipping sign (as in Fig. \ref{fig:2ls_traj}). It should be noted that the amplitude of these oscillations does change with the gradual changes to the populations.

Lasting changes to the populations do occur, however, as a result of fluctuations in the wait times between jumps. The population evolution is equivalent to simple harmonic motion with angular frequency $\Delta$, but with the additional feature that the direction changes at random, as in a telegraph process. In the regime where $\gamma\gg\Delta$, this means the populations are pinned close to their initial values for timescales much longer than $\sim1/\Delta$. However, since the wait times between jumps can vary, correlations between the populations and their initial values do eventually die off. It can be shown  that this decay of correlations occurs on a timescale proportional to $\gamma/\Delta^2\propto1/\tau_1$ \cite{de_gregorio_telegraph_2021}, in agreement with the expression for $\rho_{--}(t)$ given in Eq.~(\ref{eq:2LS_rho_alt}). For times $t\gg\tau_1$, the populations are equally likely to take on any value, as suggested by the inset of Fig. \ref{fig:2ls_traj}. This effect can also be seen if the two-time correlation function of the Pauli-Z operator is derived directly in this basis for the two-level quantum system under consideration. This correlation function can be calculated using the quantum regression theorem \cite{gardiner-book}, and it takes on the very simple form,
\be 
    \langle \sigma_z(t)\sigma_z(0)\rangle = e^{-\frac{\Delta^2}{4\gamma}t}.
\ee
Thus, the maximally mixed state is reached in the long-time limit, explaining the long-time behaviour of the state regardless of the basis in which one chooses to study its evolution. The metastable state in the energy eigenbasis, characterized by intermediate values of both the populations and coherence, can now be understood as the basis-transformed version of the metastable state we have characterized here, where the coherences have died off quickly but the populations remains close to their initial value. 
%
\section{Dissipative dynamics between distinct nearly degenerate subspaces}\label{sec:V}

\begin{figure*}
    \centering
    \includegraphics[width=0.8\textwidth, trim=50 10 50 10]{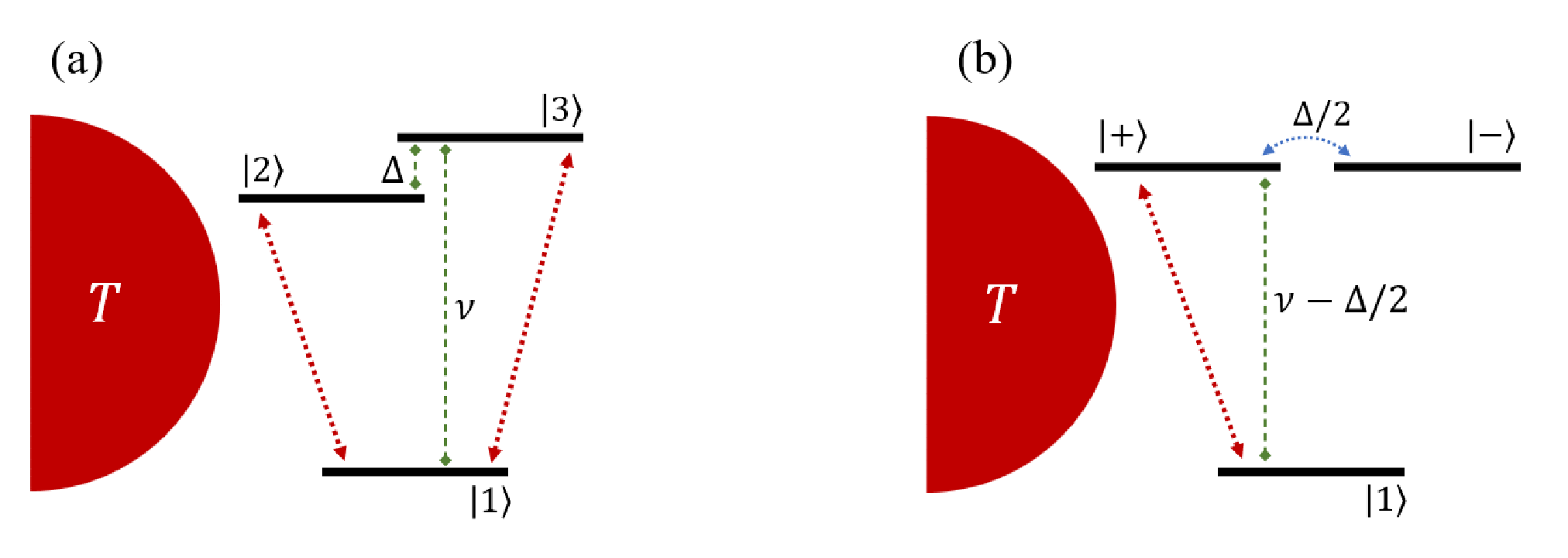}
    \caption{(a) The V model represented in the energy eigenbasis, with states and energy splittings labelled. The bath excites transitions between the ground state and both excited states (red dotted arrows). (b) The V model level diagram in the alternate basis. The bath only couples the ground state with $|+\ra$ (red dotted arrow), but $|+\ra$ and $|-\ra$ are coupled coherently with a tunnelling amplitude $\Delta/2$ (blue dotted arrow).}
    \label{fig:V_diagram}
\end{figure*}

\begin{figure*}
    \centering
    \includegraphics[width=0.95\textwidth, trim=20 10 20 10]{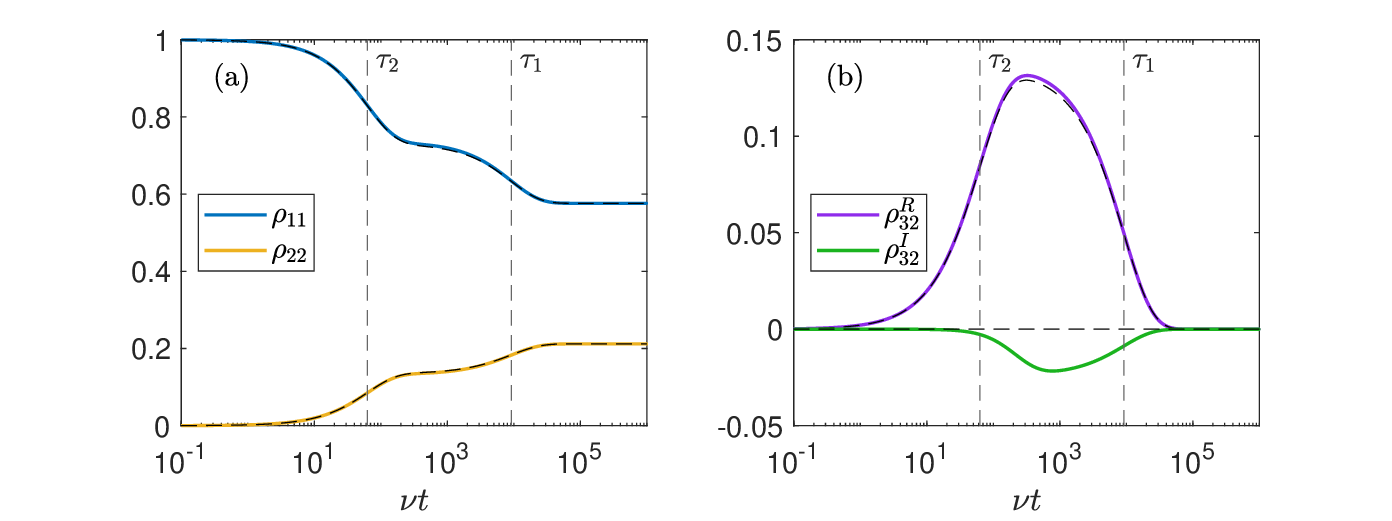}
    \caption{(a) The populations of states $|1\ra$ and $|2\ra$ and (b) the real and imaginary part of the coherence $\rho_{32}$ for the V model in the energy eigenbasis, initially in the ground state. Note that $\rho_{33}(t)=\rho_{22}(t)$ for all $t$, so only $\rho_{22}(t)$ is shown here. The dashed lines correspond to analytic perturbative calculations, showing close correspondence to the exact numerical values, with the exception of the imaginary part of the coherence which the perturbative calculations do not capture. The two characteristic timescales are shown as vertical dashed lines. $\nu=1$, $\Delta=0.001$, $a=0.02$, and $T=1$.}
    \label{fig:rho_V}
\end{figure*}

\subsection{Setup and Energy Eigenbasis Calculations}

Another situation in which dramatic timescale separation and long-lived coherences arise is that in which dissipative dynamics lead a system to jump, over finite energy splittings, into and out of distinct nearly degenerate subspaces. To investigate this situation, we consider a simple system fitting this description, the ``V" model, which consists of one ground state and two nearly degenerate excited states. This system can exhibit long-lived transient coherences similar to those identified in the two-level model of Sec. \ref{sec:2ls} \cite{tscherbul_long-lived_2014, tscherbul2015_retinal, dodin_quantum_2016}. Unlike the previous model, however, these coherences arise not only in the high-temperature limit.

The system Hamiltonian for this model is 
\be 
    H_S = (\nu-\Delta)|2\ra\la2| + \nu|3\ra\la3|,
\ee
where $\Delta\ll\nu$. The ground state (zero energy) is labelled $|1\ra$. The level diagram for this model is shown in Fig. \ref{fig:V_diagram}(a). The structure of the Hamiltonian otherwise matches Eq.~(\ref{eq:V_2ls}), but with the system part of the interaction Hamiltonian now given by
\be 
    S = \frac{1}{\sqrt{2}}\left(|1\ra\la2| + |1\ra\la3| + \text{h.c.}\right)
\ee
where h.c. represents the Hermitian conjugate.
The derivation of the unified QME for this model is analogous to that of Sec. \ref{sec:2ls}, however, we identify two distinct collapse operators:
\begin{align}
    L_{\downarrow} &= \sqrt{\frac{\gamma}{2}}|1\ra\left(\la2|+\la3|\right)
    \nonumber\\
    L_{\uparrow} &= \sqrt{\frac{e^{-\beta\nu}\gamma}{2}}\left(|2\ra+|3\ra\right)\la1|.
\end{align}
Note that these two operators represent jumps between levels separated by finite energy differences: namely, between the ground state and the coherent superposition of the two excited states given by $|+\ra=(|2\ra+|3\ra)/\sqrt{2}$. This represents a major distinction from the situation considered in Sec. \ref{sec:2ls}, as transitions in this system are associated with finite energy changes in the bath, and therefore cannot be interpreted as pure decoherence into any basis. Indeed, the presence of a finite energy splitting permits changes to the expectation value of the system Hamiltonian larger than order $\Delta$ as the system evolves. In general, the level populations of the initial state do not match those required for the Gibbs' state describing the system at equilibrium, and there {\it must} be some thermal relaxation that takes place in addition to any dephasing.

As for the two excited levels, we neglect energy differences of order $\Delta$ in defining the transition rates, allowing for the derivation of a Lindblad-form quantum master equation. We once again assume, merely for the purpose of calculations that the bath is bosonic with Ohmic spectral density, giving the downward transition rate
\begin{align}
    \gamma(\nu-\Delta,\nu)\rightarrow\gamma\equiv\gamma(\nu,\nu) =& \mathcal{J}(\nu)\left[n(\nu) + 1\right]\nonumber\\
    =& a\nu\left[n(\nu)+1\right].
\end{align}
The associated rate for an upward transition is given by $\mathcal{J}(\nu)n(\nu)=e^{-\beta\nu}\gamma$, as reflected in the expression for $L_{\uparrow}$. The presence of the finite energy splitting, $\nu$, is further reflected by the distinction between the upward and downward transition rates, whose ratio exhibits temperature dependence satisfying detailed balance relations \cite{breuer-book}. This temperature dependence will propagate through to temperature dependence of the relaxation timescales themselves.

Due to a symmetry of the system, the difference, $\rho_{22}-\rho_{33}$, between the populations of the two excited states decouples under the UQME from the rest of the dynamics and decays to zero with rate $\gamma/2$. We will consider the case where the system begins in the ground state, $\rho(0) = |1\ra\la1|$. Therefore, $\rho_{22}(0)=\rho_{33}(0)=0$, and we have that $\rho_{33}(t)=\rho_{22}(t) = (1-\rho_{11}(t))/2$ for all $t$. This allows the dynamics to reduce to a system of three homogeneous equations when we make a variable substitution, now defining $P\equiv(\rho_{33}+\rho_{22})/2 - e^{-\beta\nu}\rho_{11}$. We have,
\begin{align}\label{eq:V_UQME_eigenbasis}
    \dot{P} &= -\gamma\left(\frac{1}{2} + e^{-\beta\nu}\right)\left(P + \rho_{32}^R\right)\nonumber\\
    \dot{\rho}_{32}^R &= -\frac{\gamma}{2}\left(P + \rho_{32}^R\right) + \Delta \rho_{32}^I\nonumber\\
    \dot{\rho}_{32}^I &= -\Delta \rho_{32}^R - \frac{\gamma}{2}\rho_{32}^I.
\end{align}
Using the perturbative method outlined in Sec. \ref{sec:2ls} along with the assumption that the initial state is the ground state, we are able to obtain expressions for the ground state population and the real part of the coherence,
\begin{align}
    \rho_{11}(t) =&\nonumber\\
    \frac{1}{\mathcal{Z}(\beta)} &+ \frac{e^{-\beta\nu}}{\mathcal{Z}_I(\beta)}\left(\frac{1}{\mathcal{Z}(\beta)}e^{-\frac{\mathcal{Z(\beta)}}{\mathcal{Z}_I(\beta)}\frac{\Delta^2}{\gamma}t} + e^{-\mathcal{Z}_I(\beta)\gamma t} \right)\nonumber\\
    \rho_{32}^R(t) =& \frac{e^{-\beta\nu}}{2}\left( e^{-\frac{\mathcal{Z(\beta)}}{\mathcal{Z}_I(\beta)}\frac{\Delta^2}{\gamma}t} - e^{-\mathcal{Z}_I(\beta)\gamma t} \right),
\end{align}
where $\mathcal{Z}(\beta)=1+2e^{-\beta\nu}$ is the partition function up to leading (zero)-order in $\beta\Delta$. $\mathcal{Z}_I(\beta)=1+e^{-\beta\nu}$ is the analogous partition function for a two-level system with splitting $\nu$. As we will see, this quantity is relevant to describing the intermediate state of the system. In analogy to the two-level system, the perturbative method once again cannot capture the leading order of the evolution of the imaginary part of the coherence. We plot a selection of the matrix elements of $\rho(t)$ in Fig. \ref{fig:rho_V} as determined by solving the unified QME numerically, alongside the predictions associated with these perturbative calculations. Here we can see a good level of accuracy for the elements that the analytic method is able to capture.

Similarly to the two-level system discussed in Sec. \ref{sec:2ls}, we identify here two distinct timescales for the relaxation to equilibrium, which scale in the same manner with the transition rate, $\gamma$, and small splitting, $\Delta$. The V model, however, has a larger splitting $\nu$ which we do not assume is small relative to $T$. Therefore, the timescales depend additional on temperature via the quantity $\beta\nu$, the magnitude of which we assume nothing about. They are given by,
\begin{align}
    \tau_1 &= \frac{\mathcal{Z}_I(\beta)}{\mathcal{Z}(\beta)}\frac{\gamma}{\Delta^2}\nonumber\\
    \tau_2 &= \frac{1}{\mathcal{Z}_I(\beta)}\frac{1}{\gamma}.
\end{align}
Once again, this system exhibits long-lived coherences in the energy eigenbasis which arise after the dynamics begin, and decay to zero only on the longer timescale $\tau_1$.

In an important contrast to the two-level system, the finite splitting $\nu$ means that any bath-induced transition is necessarily accompanied by a non-negligible change in energy of the bath. Therefore, no change of basis can make the interaction with the bath ``look like" pure decoherence--the Lindblad operators cannot be simultaneously diagonalized. Accordingly, there is no basis in which each element of the density matrix evolves with only a single timescale, as we have identified for the two-level system and show in Fig. \ref{fig:rho_2ls_alternate}.

\subsection{Alternate basis and trajectory analysis}
We can, however, make an argument that is similar in spirit, by transforming to the basis, $\{|1\ra,|+\ra,|-\ra\}$, where $|\pm\ra=(|2\ra\pm|3\ra)/\sqrt{2}$. The Lindblad operators then describe jumps between basis states:
\be     L_{\downarrow}=\sqrt{\gamma}|1\ra\la+| \text{ and }L_{\uparrow}=\sqrt{e^{-\beta\nu}\gamma}|+\ra\la1|.
    \label{eq:Lpm}
\ee
Furthermore, $|+\ra$ and $|-\ra$ are not energy eigenstates, and the Hamiltonian gives rise to Rabi oscillations between them. In particular, the system Hamiltonian is
\be
    H_S = \left(\nu-\frac{\Delta}{2}\right)\left(|+\ra\la+| + |-\ra\la-|\right) + \frac{\Delta}{2}\left(|+\ra\la-|+|-\ra\la+|\right).
\ee

In this basis, the bath couples the ground state to the $|+\rangle$ level, which itself is coupled to the $|-\rangle$ site through a tunneling term $\Delta/2$. Furthermore, in this basis
it is the real part of the coherence, $\rho_{+-}^R\equiv\text{Re}[\la+|\rho|-\ra]$, that decouples from the rest of the dynamics. The UQME once again amounts to a system of three homogeneous equations. Making use of the variables $P_1 = (\rho_{++}-e^{-\beta\nu})/2$ and $P_2=(\rho_{++}-\rho_{--})/2$, where $\rho_{\pm\pm}\equiv\la\pm|\rho|\pm\ra$, these are,
\begin{align}\label{eq:uqme_V_transformed}
    \dot{P_1} &= -\gamma\mathcal{Z}_I(\beta)P_1 - \frac{\Delta}{2}\rho^I_{+-}\nonumber\\
    \dot{P_2} &= -\gamma P_1-\Delta\rho^I_{+-}\nonumber\\
    \dot{\rho}^I_{+-} &= \Delta P_2 -\frac{\gamma}{2}\rho^I_{+-}.
\end{align}

\begin{figure}
    \centering
    \includegraphics[width=0.9\columnwidth, trim=5 10 5 10]{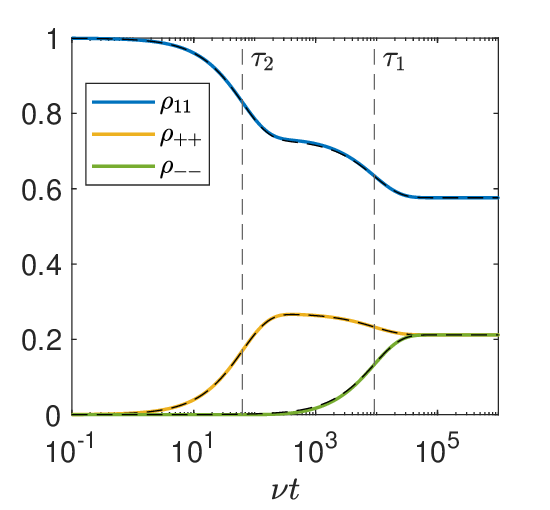}
    \caption{Populations of the three levels of the V model in the $\{|1\ra,|+\ra,|-\ra\}$-basis, starting from the ground state. The dashed curves represent the results of analytic perturbative calculations. As with the other setups considered, thermalization takes place over two distinct timescales. Here, only the slower timescale, $\tau_1$, shows up in the evolution of the population $\Tilde{\rho}_{--}$, which is not directly involved in the dissipative dynamics. The two characteristic timescales are shown as vertical dashed lines. Parameter values are the same as in Fig. \ref{fig:rho_V}.}
    \label{fig:rho_V_transformed}
\end{figure}

Using the same perturbative method to obtain expressions for the populations, $\rho_{++}$ and $\rho_{--}$, of state $|+\ra$ and $|-\ra$, respectively, we find:
\begin{align}
    \rho&_{++}(t) = \nonumber\\
    &\frac{e^{-\beta\nu}}{\mathcal{Z}(\beta)} + 
    \frac{e^{-\beta\nu}}{\mathcal{Z}_I(\beta)}
    \left( \frac{e^{-\beta\nu}}{\mathcal{Z}(\beta)}e^{-\frac{\mathcal{Z(\beta)}}{\mathcal{Z}_I(\beta)}\frac{\Delta^2}{\gamma}t}
    - e^{-\mathcal{Z}_I(\beta)\gamma t} \right)\nonumber\\
    \rho&_{--}(t) = \frac{e^{-\beta\nu}}{\mathcal{Z}(\beta)}\left(1-e^{-\frac{\mathcal{Z(\beta)}}{\mathcal{Z}_I(\beta)}\frac{\Delta^2}{\gamma}t}\right).
\end{align}
The perturbative method does not capture the dynamics of $\rho^I_{+-}$--luckily, the populations in this basis are sufficient to draw the insights we want to discuss. We plot in Fig. \ref{fig:rho_V_transformed} the populations obtained by solving the master equation numerically, alongside curves representing these expressions. Transforming the results plotted in Fig. \ref{fig:rho_V_transformed} back into the energy eigenbasis yields behaviour very close to the results of the direct eigenbasis calculations shown in Fig. \ref{fig:rho_V}.

\begin{figure*}
    \centering
    \includegraphics[width=0.8\textwidth, trim = 10 0 10 10]{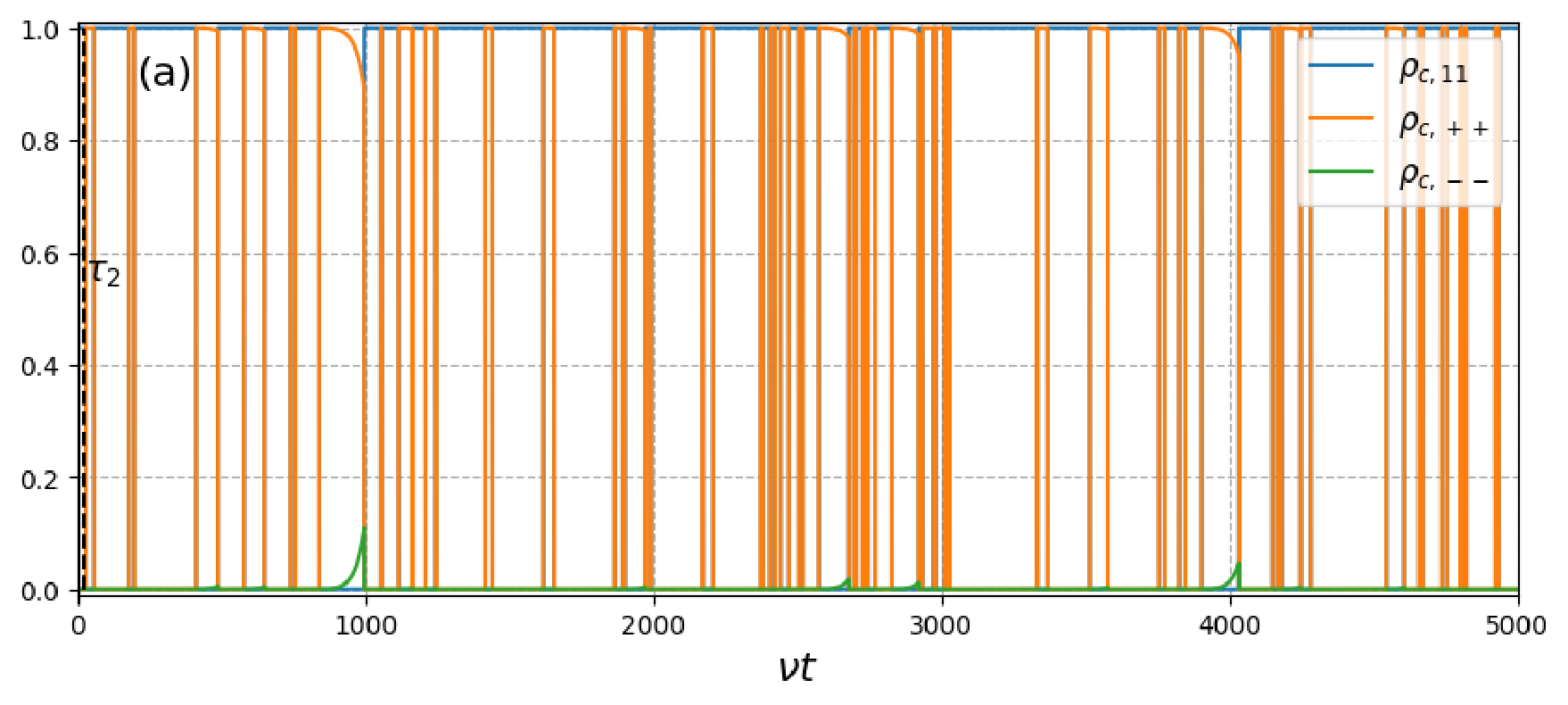}
    \includegraphics[width=0.8\textwidth, trim = 17 10 15 0]{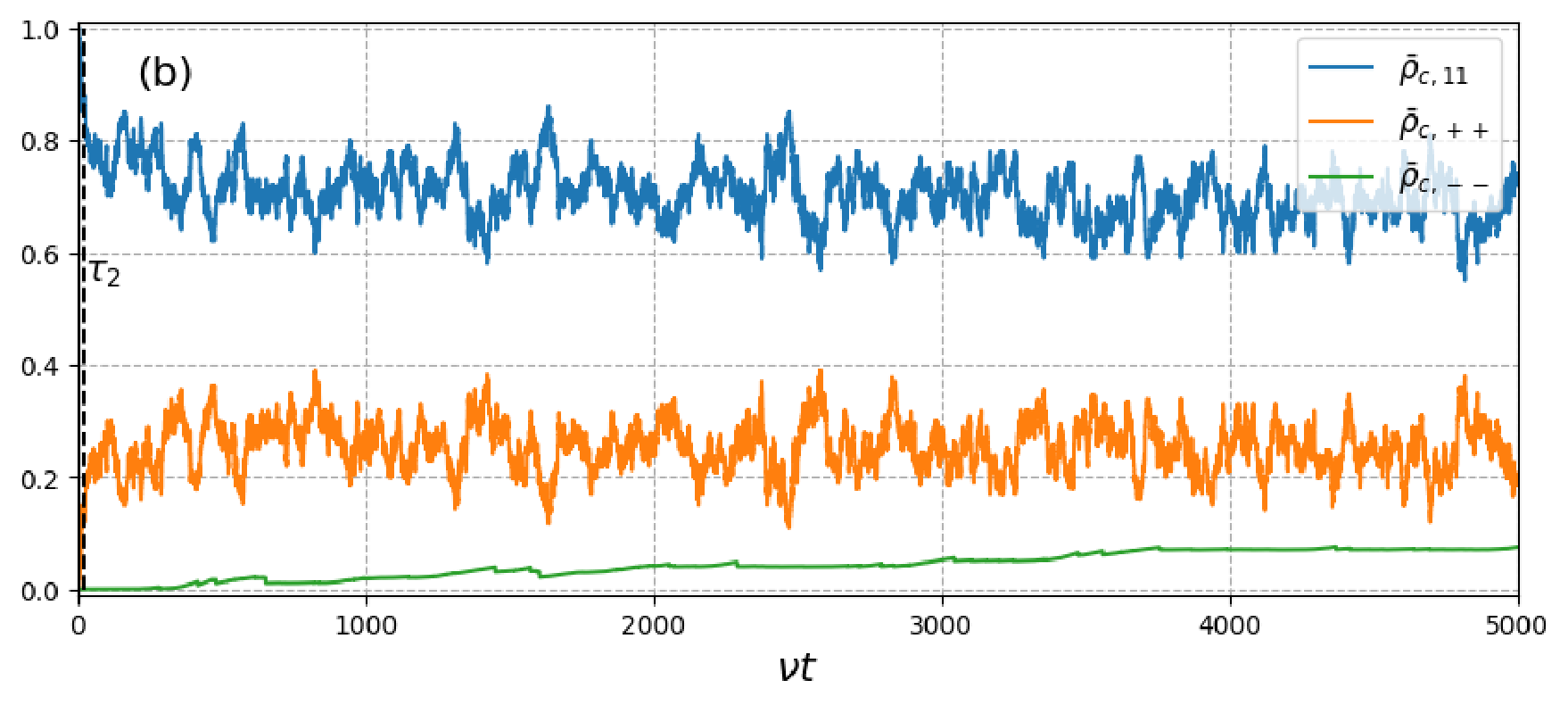}
    \caption{(a) An example trajectory for the V model in the $\{|1\ra,|+\ra,|-\ra\}$-basis, showing early stages of the dynamics only. A bath-induced jump instantaneously sends the population of $|+\ra$ (orange) to 1 as the population of $|1\ra$ (blue) goes to 0, or vice-versa. Only once the system has jumped to state $|+\ra$ can any population reach $|-\ra$ through Rabi oscillations. Typically, since $\Delta\ll\nu$, this process is interrupted by a jump back down to $|1\ra$ well before a single period of the Rabi oscillations is completed. (b) one hundred such trajectories, averaged. This averaging recovers the density operator given by quantum master equations, with noise due to the limited size of the sample. Parameter values are the same as in Fig. \ref{fig:rho_V}.}
    \label{fig:V_transformed_traj}
\end{figure*}

Interestingly, only the longer timescale $\tau_1$ appears in the expression for $\rho_{--}$. This can be understood as a result of the fact that $|-\ra$ is isolated as the state that is not ``included" in the dissipative dynamics, see Eq. (\ref{eq:Lpm}). Bath-induced jumps from the ground state, which occur with a rate $e^{-\beta\nu}\gamma$, serve only to populate state $|+\ra$ directly. Population can only arrive in $|-\ra$ via the Rabi oscillations that take place after a jump up to $|+\ra$. The period for these Rabi oscillations is proportional to $1/\Delta$, but the timescale for $\rho_{--}$ to reach its steady-state population is increased to $\sim\Delta^2/\gamma$ by the fact that these Rabi oscillations are often quickly interrupted by a jump back down to the ground state, as shown in Fig. \ref{fig:V_transformed_traj}(a). Figure \ref{fig:V_transformed_traj}(b) displays an average of 50 distinct trajectories of the type depicted in Fig. \ref{fig:V_transformed_traj}(a), showing that averaging as such indeed recovers the non-conditional reduced density operator. Interestingly, the averaged trajectories for $\rho_{11}$ and $\rho_{++}$ exhibit far more noise than that for $\rho_{--}$, likely attributed to the special status of $|-\ra$ as the state not directly involved in the dissipative dynamics, and thus not involved in the frequent jumps that occur as a consequence of the coupling to the bath.

The approach to equilibrium can be understood to happen in two distinct phases. The first is thermalization with respect to the larger energy gap, $\nu$, which occurs over a shorter timescale $\tau_2$ set by the rate constant, $\gamma$, characterizing the dissipative dynamics. The second is the ``mixing" of the two excited states $|+\ra$ and $|-\ra$, which happens over slower timescale due to the constraints placed on the dynamics by the Hamiltonian. We can understand this by considering how we neglect the energy difference $\Delta$ between states $|2\ra$ and $|3\ra$ in describing the dissipative dynamics. Because of this, $\rho_{22}=\rho_{33}$ in the Gibbs state describing the system at equilibrium, i.e., the steady state is ``maximally mixed" with respect to these two states. Thus, $2\times2$ sub-matrix of $\rho$ describing the excited states and their coherences is proportional to $I_2$ at equilibrium, and invariant under any change of basis that affects only these two states (like the one we have used).

The thermalization dynamics must therefore reflect both that (i) at steady state, $\rho_{++}=\rho_{--}=e^{-\beta\nu}\rho_{11}$, and that (ii) only the state $|+\ra$ is accessible to the system on the shorter timescale $\tau_2$ over which thermalization with respect to the larger splitting takes place. Indeed, at intermediate times $\tau_2\ll t\ll\tau_1$, the populations of $|1\ra$ and $|+\ra$ are
\begin{equation}
    \rho_{11} \approx \frac{1}{\mathcal{Z}_I(\beta)}\text{ and } \rho_{++}\approx \frac{e^{-\beta\nu}}{\mathcal{Z}_I(\beta)},
\end{equation}
which is simply the Boltzmann distribution for a two-level system. This suggests that the first stage consists of the system thermalizing as if it were a two-level system, with the system later proceeding towards its ``correct" equilibrium state, once sufficient time has passed for the Rabi oscillations to have an effect on the density operator.



\section{Discussion}\label{sec:summary}
We have investigated two situations in which open quantum systems with nearly degenerate eigenstates exhibit a dramatic separation of timescales, and associated long-lived coherences, in the process of relaxation towards equilibrium.
These effects are captured when dynamics are modelled using the unified quantum master equation, which is better suited to such systems than the fully secular master equation. In examples, we have considered quantum systems interacting with bosonic baths with ohmic spectral density. However, these choices were made only for the purposes of calculating transition rates, and are not essential to the observation or explanation of the phenomena under investigation. Our approach towards gaining insight as to the physics underlying these timescale effects has been two-fold.

Firstly, we have considered how transforming to different bases can reveal different characteristics of the time evolution. In particular, when the system part of the system-bath interaction Hamiltonian takes on a nontrivial, non-diagonal form, the energy eigenbasis may not be the only natural basis in which to study the dynamics. Transforming to a basis that simplifies the interaction Hamiltonian and/or Lindblad operators can lead to simplified expressions for the state as a function of time and aid in the analysis.

Secondly, using a QME that is of Lindblad form has allowed us to straightforwardly simulate quantum trajectories consistent with these dynamics. While the reduced density operator takes into account the presence or absence of phase coherence between separate instantiations of a quantum system interacting with a bath, individual trajectories, in a sense, provide a more insightful visualization of the underlying dynamics, with randomly timed jumps punctuating periods of coherent evolution.

The presence of nearly degenerate levels in an open quantum system means that, up to leading order in the small splitting between them, which we have denoted $\Delta$, the reduced density operator at equilibrium is maximally mixed with respect to the nearly degenerate subspace. As such, it will be invariant under any basis transformation that only affects states within this subspace, mimicking the ``basis freedom" that occurs for systems with strict degeneracies. For this reason, while the system Hamiltonian technically does specify a unique energy eigenbasis, using a basis chosen by considering the system-bath interaction leads to a clearer understanding of the distinct timescales that arise in the evolution. This was seen very directly in Sec. \ref{sec:2ls}, where it turned out that the interaction between our two-level system and its bath could be interpreted as simple decoherence, but into a basis other than the energy eigenbasis. The long-lived coherence, therefore, is highly analogous to spurious coherences that might arise when investigating quantum systems with strict degeneracies, but which are uncontroversially explained away in these scenarios with an appeal to the well understood basis freedom.

In Sec. \ref{sec:V}, we considered the case where dissipative dynamics excite jumps over finite energy splittings between distinct nearly degenerate subspaces, giving rise to a set of non-simultaneously diagonalizable Lindblad operators. The system-bath interaction could not, therefore, be interpreted as pure decoherence in any basis. However, the same type of basis freedom was employed to identify a basis more natural for describing the dynamics, without having any effect on the reduced density operator at equilibrium. This amounted to identifying, within the nearly degenerate subspace, a pair of orthogonal states, of which one is directly involved in the dissipative dynamics and one is not. Once again, this led to some clarity surrounding the existence of distinct timescales in the dynamics, as it elucidated the fact that the shorter timescale is associated more closely with the dissipative part of the dynamics only, while the longer timescale characterizes effects that involve an interplay between the coherent dynamics (Rabi oscillations) and the dissipative dynamics.

For both cases considered, changes of basis lent themselves well to the inspection of quantum trajectories, as differences between how different matrix elements evolve at the trajectory level, both qualitatively and quantitatively, reflect the different timescales that show up in the expressions for the non-conditional reduced density matrix elements. Certain open questions do, however, remain. For instance, while it is immediately evident by inspecting the trajectories for the two-level system in Fig. \ref{fig:2ls_traj} how an average of many statistically similar trajectories would recover the curves in Fig. \ref{fig:rho_2ls_alternate}, the same cannot quite be said of those for the V model in Fig. \ref{fig:V_transformed_traj}. Looking at a single trajectory, it appears that the population of $|-\ra$ only very rarely has a chance to increase beyond about 0.01 or 0.02 before a jump occurs, resetting it to zero. It therefore seems unintuitive that at steady state, this population gets as high as $e^{-\beta\nu}/\mathcal{Z}(\beta)$ (about 0.21 for the parameter values plotted). This is not to dispute the results of the calculations, just to emphasize the limitations of building intuition by studying an individual trajectory in a particular basis while the reduced density operator is really an ensemble average of infinitely many.

We wish to distinguish these results from related work investigating the presence of coherence at steady state in nonequilibrium situations for open quantum systems that include near degeneracies \cite{guarnieri2018, purkayastha_tunable_2020, ivander_quantum_2022}. Employing the framework of quantum trajectories to gain insight into the physical processes underlying these steady-state coherences remains a open direction for future research. Furthermore, it is worth noting that recent studies have worked towards establishing methods of obtaining quantum trajectories even if the dynamics are described by non-GKLS-form master equations \cite{donvil2022,becker_quantum_2023}. Such methods would, in principle, allow analyses like this for systems even when no GKLS-form master equation serves as a valid approximation for the dynamics.


\begin{acknowledgements}
D.S. acknowledges the NSERC discovery grant and the Canada Research Chairs Program.
M.G. acknowledges support from the NSERC Canada Graduate Scholarship-Doctoral. 
M.J.K. acknowledges the financial support from a Marie Sk\l odwoska-Curie Fellowship (Grant Agreement No.~101065974) and the Centre for Quantum Information and Quantum Control (CQIQC) at the University of Toronto (UofT) for partially supporting his visit to the UofT during which this project was initiated. 
\end{acknowledgements}

\bibliography{ref}
\end{document}